\documentstyle[preprint,aps,epsf]{revtex}

\def\be{\begin{equation}}
\def\ee{\end{equation}}

\def\o{\over}
\def\ba{\begin{array}}
\def\ea{\end{array}}
\def\bea{\begin{eqnarray}}
\def\eea{\end{eqnarray}}

\begin{document}
\title{Stable bimaximal neutrino mixing pattern}
\author{Gautam Dutta\footnote{email: gautam@mri.ernet.in}}
\maketitle
\begin{abstract}
The neutrino oscillation experiments increasingly point towards a mixing
pattern that can be parametrised with two near maximal and one small mixing
angle. We investigate whether such a mixing pattern can be generated as a 
fixed point of renormalisation group evolution.
\end{abstract}

\section{Introduction}
The data from the various neutrino experiments indicate a bimaximal mixing
among three neutrino generations. 
The atmospheric neutrino puzzle  
requires a near maximal 
mixing between a pair of neutrinos \cite{atmos}.
The solar neutrino puzzle has various possible solutions but the large angle
MSW solution is favoured \cite{bahcall}. As the mass squared difference required for these
two solutions differ by about few orders of magnitude, we must have two
different pairs with large mixing. 
The CHOOZ experiment constrain one of the mixing angles to be 
small \cite{chooz}. So out of three mixing angles two are large and one is small
indicating a bimaximal structure.    
The large mixing in neutrino is in sharp contrast
to the very small mixing in the quark sector. Flavor symmetries 
may generate bimaximal mixing structure in neutrino masses \cite{bimax}. 
But such
structure generated at the high scale may not be stable under
renormalisation group evolution to low scale if neutrino masses are
quasi degenerate \cite{ma,chankowski}. Alternatively
renormalisation Group evolution from high scale to the present scale have been
proposed as a mechanism to generate large mixing from a small mixing
\cite{balaji}. But in these mechanism enhancement of the mixing angle is
scale dependent and hence may be unstable and fine tuned. \cite{dutta1}. 

We do a three generation analysis and find that large mixing can be a stable
fixed point of renormalisation group evolution in certain region of
experimentally allowed parameter space. 
We consider a quasi degenerate mass structure with one of the state in an
opposite CP phase to the other two. 

Existence of large mixing angle as a
stable fixed point of renormalisation group evolution has the advantage that
once the mixing attains the stable large value, it doesn't change much
under renormalisation group evolution and hence there is no fine tuning. So
one doesn't have to explicitly evolve the mixing matrix under
renormalisation group. The fixed point value of the mixing can be obtained
after a small perturbation on the neutrino mass matrix at the high scale 
due to radiative correction.   

One may take the view that unlike the quark and charged lepton mixing, all
the three neutrino mixing angle can be large at the high scale. 
This is because quark and charged lepton
masses show a hierarchical structure while neutrinos may not. In this case
we investigate whether one of the mixing angle attain small value as fixed 
point due to radiative correction while the other two remain unchanged. If so
then this mechanism can also produce the required bimaximal mixing pattern. 
In the next section we present the analysis to generate large stable mixing
due to perturbation from radiative correction. In section 3 we investigate
whether a stable small mixing can be obtained from all large mixing at high
scale. In section 4 we present the conclusion. 

\section{Radiative enhancement of mixing}
Consider a Majorana mass structure with three almost degenerate masses. 
The CP phase of one of the species is opposite to that of the other
two.
\be
M_D=(m,m,-m)     \label{md1}
\ee
Such a degenerate mass pattern can be obtained from a seesaw mechanism 
\cite{seesaw} with 
non-diagonal structure in right handed Majorana masses\cite{dutta}. In these
models a pair of opposite CP states are maximally mixed and they form the 
pseudo Dirac neutrinos \cite{wolfenstein}. 
The other two mixing angles are small. 
Let us parametrise the mixing matrix $U$ as follows
\be
U(\psi,\phi,\omega) = R_{23}(\psi)R_{13}(\phi)R_{12}(\omega) \label{mix}
\ee
The mixing matrix can contain two physical Majorana phases. We ignore these
phases in our calculations. A detailed analysis including these phases and
different types of hierarchy amongst the masses have
been done by Haba {\it et. al.} \cite{haba1} in MSSM. The interesting cases
of the stability of maximal and zero mixing occurs when the difference of
these physical phases is 0 or $\pi/2$ and all the masses are degenerate. We
do a model independent analysis through a perturbation of the mass matrix at
the high scale and investigate whether a bimaximal mixing can be a fixed
point of the evolution of the mass matrix under radiative correction.     

Let the (2,3) states be the pseudo Dirac pair in $M_D$. So $\psi$ is maximal
and $\phi$ and $\omega$ are small.   
In this parametrisation and the mass spectrum eq.(\ref{md1}), the mass
matrix in the flavor basis will be
\bea
M&=&UM_DU^{\dagger} \\
 &=& m\left(\ba{lll} 
     C_{2\phi}&-S_{2\phi}S_\psi & -S_{2\phi} C_\psi \\
     -S_{2\phi}S_\psi & (1-2C^2_\phi S^2_\psi) & -S_{2\psi} C^2_\phi \\
     -S_{2\phi} C_\psi & -S_{2\psi} C^2_\phi & 1-2C^2_\phi C^2_\psi 
     \ea \right )
\eea
Due to radiative correction the effective mass matrix at the low scale $M_Z$
is \cite{babu,ellis,chankowski1}
\be 
M(M_Z)=\left(\ba{ccc} 1+\delta_e & 0 & 0 \\ 0 & 1+\delta_\mu & 0 \\
         0 & 0 & 1+\delta_\tau \ea \right)  M
       \left(\ba{ccc} 1+\delta_e & 0 & 0 \\ 0 & 1+\delta_\mu & 0 \\
         0 & 0 & 1+\delta_\tau \ea \right)      \label{rad}
\ee    
Let 
\[
\delta_\mu=\delta_e + \delta_{e\mu} {\hbox{ and}} 
\delta_\tau=\delta_e + \delta_{e\tau}
\]
   
To the first order in $\delta_e, \delta_{e\mu}$ and $\delta_{e\tau}$, 
$M(M_Z)$ can be written as 
\be 
M(M_Z) = (1+2\delta_e)M + \delta M   \label{mmz}    
\ee
where
\be
\delta M= m\left(\ba{lll}
      0 &-S_{2\phi}S_\psi \delta_{e\mu} & 
        -S_{2\phi} C_\psi\delta_{e\tau} \\
     -S_{2\phi}S_\psi \delta_{e\mu} & 
     (1-2C^2_\phi S^2_\psi)2\delta_{e\mu} & 
     -S_{2\psi} C^2_\phi (\delta_{e\mu}+\delta{e\tau})\\
     -S_{2\phi} C_\psi \delta_{e\tau} & 
     -S_{2\psi} C^2_\phi (\delta_{e\mu}+\delta{e\tau}) & 
      1-2C^2_\phi C^2_\psi 2\delta_{e\tau}    
     \ea \right )
\ee
Here we see that due to degeneracy in the (1,2) sector, and the suitable
parametrisation the angle $\omega$ doesn't enter the mass matrix $M$. So
$\omega$ can be arbitrary at the high scale. 

The first part of $M(M_Z)$ in (\ref{mmz}) $(1+2\delta_e) M$ is diagonalised
by the same mixing matrix $U(\psi,\phi,\omega)$. The three eigenvalues are 
just $1+2\delta_e$ times the eigenvalues of $M$. So the degeneracy is not
lifted in this part. Treating the second term $\delta M$ as a perturbation
over $(1+\delta_e)M$ we get the following split in the degenerate mass
eigenvalues
\be
\delta m_{1,2}=(M_{22}\pm\sqrt{M_{22}^2+M_{12}^2})\delta_{e\mu}     
                                                         \label{ev2}
\ee
The corresponding mass eigenvectors in this sector are given by
a rotation in the (1,2) sector $R_{12}(\omega')$ where $\omega'$ is given by  
\be
\tan\omega' ={M_{12}\o M_{22}+\sqrt{M_{22}^2 + M_{12}^2}}
              \label{omg'}  
\ee
The changes in the state 3 is small and from the parametrisation of the 
mixing matrix in eq.(\ref{mix}) we see that the angles $\psi$ and $\phi$ do
not change much. So the mixing after the perturbation is given by
\be
U(\omega',\psi',\phi') = R_{23}(\psi')R_{13}(\phi')R_{12}(\omega')
                                                           \label{mix'}
\ee
where $\psi'\approx \psi$ and $\phi' \approx \phi$. 

Eq.(\ref{omg'}) says that $\omega'$ is independent of radiative corrections
($\delta_e,\delta_\mu$ and $\delta_\tau$) at a given scale. 
So $\omega'$ achieves
the stable value independent of renormalisation Group Evolution.

\begin{figure}[p]
\centerline{
\epsfxsize=10cm \epsfysize=15cm
\epsfbox{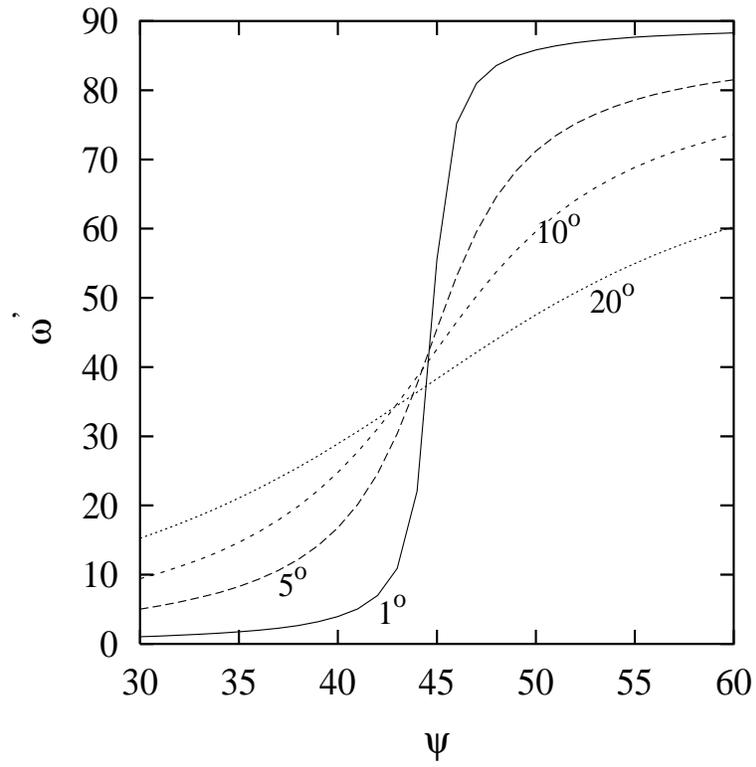}
           }
\caption{Fixed point values of $\omega'$ as a function of $\psi$ for various 
values of $\phi = 1^o, 5^o, 10^o$ and $20^o$.} 

\label{omgpsi}
\end{figure}

The stable value of $\omega'$ depends on the high scale values of $\phi$ and
$\psi$.  
From eq.(\ref{omg'}) we see that when $M_{22} << M_{12}$ $\omega'$ is near
maximal. This happens for small $\phi$ and near maximal $\psi$.
In figure \ref{omgpsi} we plot $\omega'$ verses $\psi$ for various
values of $\phi$.

The interesting point is that large $\omega'$ is obtained for small values of 
$\phi$ and very near maximal value of $\psi$.  This is the region of interest 
suggested by the CHOOZ constraint and the atmospheric neutrino data. 
If the high scale mass matrix is generated by a seesaw mechanism that
produces degenerate neutrino masses with a pseudo Dirac pair, it is natural
to have $\phi$ and $\omega$ small at the high scale while $\psi$ is maximal
\cite{dutta}. 
Then due to quantum correction $\omega$ stabilises to a large value by
radiative correction. This mechanism is model independent. It will work in
any theory where the radiative corrections can be expressed as in
eq.(\ref{rad}). 
The drawback of this mechanism is that the angle $\psi$ is constrained to be 
in a very small window around
$45^o$ for small values of $\phi$ if we want large $\omega'$. For example if
we demand $\omega$ to be above $30^o$ then $\psi$ has to be between $44^o$
and $45^o$ for $\phi=1^o$. For $\phi=10^o$ the region of $psi$ is
constrained to be between $42^o-45^o$. If $\psi$ lies below $40^o$ this
mechanism cannot explain the large angle solar neutrino solution. However it
may be possible that $\psi$ may be almost exactly maximal as suggested by
the current atmospheric neutrino data. 
To make a simple estimate about how much $\psi$ can deviate from maximal
mixing due to radiative correction, we evaluate the small perturbative change 
in the third eigenstate with $CP=-1$. As we are interested in small $\phi$
region we put $\phi=0$. So after radiative correction the third mass state
with eigenvalue $-m$ is 
\be
|3'\rangle = \left(\ba{c} 0 \\ S_\psi \\ C_\psi \ea \right) +
       {S_{2\psi}C_\phi^2S_{\omega'}(\delta_{e\mu}+\delta_{e\tau})\o 2}
       \left(\ba{c} 0 \\ C_\psi \\ -S_\psi \ea \right)
\ee
This shows that the change in the state $|3\rangle $ indicates a change
in the angle $\psi$ by ${S_{2\psi} S_{\omega'}\delta_\tau \o 2}$ in radians, 
i.e $({S_{2\psi} S_{\omega'}\delta_\tau \o 2}{180 \o \pi})^o$. 
If $\delta_\tau \approx 10^{-2}$ then
the change in the angle $\psi$ is less than $0.3^o$ around maximal $\psi$. 
So the angle $\psi$ seems
to be stable enough to stay near maximal within few degrees. 
   
\section{A variant of the mechanism}
In the above section we considered generation of a large mixing from a small
mixing at the high scale. However we may have the view that having large
mixing between the mass states can be as natural as having small mixing. Then
radiative correction may produce a small mixing between a pair with same CP
parity while not affecting the other two large mixing. Then this can be an
alternative mechanism to generate a stable bimaximal mixing pattern. Here we
see if this can be done. 

We consider the following Majorana mass spectrum at the high scale 
which is different from
eq.(\ref{md1}).
\be
M_D=(m,-m,m)     \label{md2}
\ee
Let us parametrise the mixing matrix $U$ as follows
\be
U(\omega,\psi,\phi) = R_{23}(\psi)R_{12}(\omega)R_{13}(\phi) \label{mix2}
\ee

Note that this parametrisation of mixing matrix is different from that in
the standard parametrisation given in eq.(\ref{mix}). This parametrisation
is useful here as we wish to find the stable value of the (1,3) mixing angle
$\phi$ after radiative correction as the these two states have the same CP
parity. As long as $\phi$ is small this parametrisation will have nearly 
the same
prediction for the mixing angles $\omega$ and $\psi$ as with the standard
parametrisation in eq.(\ref{mix}). This is because for small values of
$\phi$, $R_{13}(\phi)$ is very nearly identity and hence it commutes with
all other matrices. Hence its position doesn't change the predictions as
long as the relative order of the large mixing matrices $R_{12}(\omega)$ and
$R_{23}(\psi)$ are not changed. 

The mass matrix in the flavor basis is
\bea
M&=&UM_DU^{\dagger} \\
 &=& m\left(\ba{lll} 
     C_{2\omega}&-S_{2\omega}C_\psi & S_{2\omega} S_\psi \\
     S_{2\omega}C_\psi & (1-2C^2_\omega C^2_\psi) & S_{2\psi} C^2_\omega \\
     S_2\omega S_\psi & S_{2\psi} C^2_\omega & 1-2C^2_\omega S^2_\omega 
     \ea \right )                                   \label{Mflavor}
\eea
Note that the angle $\phi$ doesn't appear in $M$ due to the degeneracy in
the (1,3) states.

With this parametrisation and the radiative correction given by eq.(\ref{rad})
the stable value of the mixing angle $\phi$ is 

\be
\tan\phi' = {M_{13} \o M_{33}+\sqrt{M^2_{33} + M^2_{13}} }   \label{tanphi}
\ee
 
\begin{figure}[p]
\centerline{
\epsfxsize=10cm \epsfysize=15cm
\epsfbox{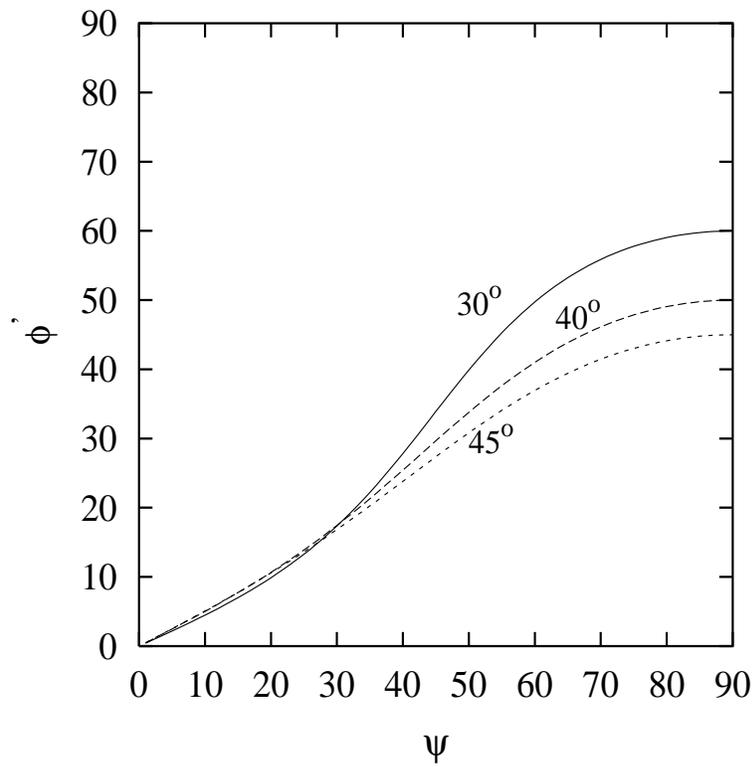}
           }
\caption{Fixed point values of $\phi'$ as a function of $\psi$ for various
values of $\omega = 30^o, 40^o$ and $45^o$.}

\label{phipsi}
\end{figure}

We see from eq.(\ref{Mflavor}) that $\phi'$ is not small unless $\omega$ and
$\psi$ both are small. In fig\ref{phipsi} we show a plot of the fixed point
value of $\phi$ as a function of $\psi$ for various values of $\omega$. For
large values of $\omega(>30^o)$ we see that $\phi$ is not small near large
or maximal $\psi$.  So we cannot produce a bimaximal mixing in this way.  

\section{Conclusion}
Bimaximal mixing pattern can be obtained from small mixing at high scale. We
present an analysis based on finding stable fixed point of mixing angle
after small perturbation to the mass matrix due to radiative correction.
This has the advantage that the large stable mixing can be evaluated without
requiring to evolve the mixing from high scale down to the present scale.
This is possible because the mixing doesn't change with scale further down 
after attaining the fixed point value. The large mixing being a fixed point
this solution is not fine tuned as is the case with some of the earlier
analysis. We present a variant of the mechanism to see whether one small
mixing can be obtained from all large mixing at high scale by radiative
correction. We find that this is not possible in the region of interest
indicated by experiments.    

\end{document}